\documentclass[a4paper, 11pt]{article}
\usepackage[utf8]{inputenc}
\usepackage{color}
\usepackage{comment} 
\usepackage{lipsum} 
\usepackage{fullpage} 
\usepackage{graphicx} 

\usepackage[
    natbib=true,
    style=numeric,
    sorting=none
]{biblatex}

\graphicspath{ {images_color/} }

\begin{document}

\title{Heterogeneity promotes first to second order phase transition on flocking systems }

\author{Leandro Guisandez\footnote{lguisandez@iflysib.unlp.edu.ar}, Gabriel Baglietto and Alejandro Rozenfeld}

\maketitle

\begin{abstract}

We have considered a variation of the Vicsek model with vectorial noise where each one of the agents have their own noise amplitude normally distributed around a mean value, $\mu$, with standard deviation $\sigma$. First-order phase transition are observed for standard deviation $0\leq\sigma<\sigma_{tri}\approx0.11$, whereas for larger values, up to $\sigma=0.3$, a continuous phase transition occurs. 
For values of $\sigma$ in the interval $0.15\leq\sigma\leq0.30$ the continuous nature of the observed transition is characterized by means of finite-size scaling techniques, that also allow us to estimate the exponents driving the transition. A study of bands stability suggests that no band can form in this regime. Inspired by biological facts, the perception heterogeneity introduced in the model trough $\sigma$, allow us to tune the collective behaviour of the system. 

\end{abstract}



\section{Introduction}

The study of active matter has attracted attention in last years~\cite{menzel2015tuned,vicsek2012collective,marchetti2013hydrodynamics,ramaswamy2010mechanics}. 
Numerous examples are found in the literature from living systems like bacterial colonies~\cite{ben1994generic,peruani2012collective}, cells tissue~\cite{szabo2006phase}, insect swarms~\cite{buhl2006disorder}, flocks, fish schools~\cite{becco2006experimental} to non-living self-propelled particles like rods~\cite{narayan2007long,kudrolli2008swarming}, disks~\cite{deseigne2010collective}, and Janus particles~\cite{walther2008janus}. 
An exciting aspect of many of these systems is the existence of a phase transition between a gas-like state, where agents move randomly, and a state of collective motion, where they may form flocks, bands~\cite{gregoire2004onset} or even more complex patterns~\cite{strombom2011collective}.
From the modelling point of view, the nature of the phase transition suffered by the system is a relevant aspect as different type of phase transition are associated with different groups of phenomena.
While first-order phase transitions are characterized by an abrupt change in the state of the system, hysteresis and phase coexistence~\cite{binder1997applications}, second-order phase transition display long range correlations (at the transition point, known as critical point), universality, anomalous fluctuations and high susceptibilities to external fields~\cite{binder1997applications}.

As it happens for most complex systems the macroscopic dynamics of these active matter ensembles, including the nature of the phase transition they eventually suffer, depends on the way the individuals interact each other, the type of connectivity and the features of the individuals themselves. 
Much attention has been payed to the first two of these factors finding that while minimal metric models for flocking tend to exhibit first-order phase  transitions~\cite{gregoire2004onset,chate2008modeling,chate2008collective,baglietto2013gregarious}, their topological counterparts show continuous phase transitions and features of criticality~\cite{ginelli2010relevance,barberis2014evidence}, the latter also observed in actual flocks~\cite{Ballerini1232,cavagna2010scale,attanasi2014finite,chen2012scale,peruani2012collective}. 
On the other hand, from a biological point of view, the features of the individuals themselves, like phenotype and physiological traits, are thought to be relevant factors that shape the interaction among individuals~\cite{seebacher2017physiological} since each of these individuals reacts in its own manner for the same stimulus~\cite{maye2007order}. 
In other words, a common feature of biological systems, heterogeneity, plays an important role in the collective behaviour of the system~\cite{von2015solving}. 
In fact, collective motion has been proved to be sensitive to the presence of agents with different capabilities in self-propelled particles systems. 
In a mixture of aligners and non-aligners collective behaviour strongly depends on the density of non-aligners in such a way that can be completely smeared~\cite{copenhagen2016self,yllanes2017many}. 
Binary mixtures~\cite{ariel2015order,menzel2012collective}, presence of leaders~\cite{ferdinandy2016collective} and particles with variable velocities~\cite{mishra2012collective} are other examples where collective motion is striked by multi-agent component. 
Even hierarchical organization arises when the interaction is not the same for all individuals~\cite{zamani2017glassy}.
~\citet{lafuerza2013effect} have observed that, for interacting particles jumping between two states with different transition rates, the fluctuation of global variables of the system may increase or decrease depending on the degree of heterogeneity, showing once again that the combination of heterogeneity and stochasticity results in unexpected and non-trivial behaviour. 

A question that, however, remains open is how heterogeneity affects the onset of collective motion. 
The physics of disordered systems in equilibrium gives us an inspiring starting point for answer this question. 
In these well studied systems it is known that introducing heterogeneity may alter the nature of the suffered phase transition or even its universality class~\cite{imry1979influence,aizenman1990rounding,cardy1999quenched,chatelain2001softening}. 
Therefore we propose to study the nature of the phase transitions appearing in flocking systems introducing in the Vicsek model (VM)~\cite{vicsek1995novel} a heterogeneity inspired by biological systems.
The original VM considers identical self-propelled agents interacting through a simple alignment rule perturbed by noise and exhibit the most studied phase transition in a flocking system. 
Here we add heterogeneity to the VM by allowing each individual to have its own noise amplitude taken from a given distribution. In a biological context this means introducing a phenotype for each individual through an effective factor, the noise amplitude. For populations of agents with noises normally distributed around a mean value, $\mu$, with a standard deviation $\sigma$, we observe for the very first time in flocking systems a complex scenario with both first-order and continuous phase transitions depending on the degree of heterogeneity of the sample. 

\section{Model and Simulation setup}\label{model}

We consider a system of $N$ interacting particles moving with constant speed $v_0$ in a two-dimensional rectangular box of size $L=\sqrt{N}$ with periodical boundary conditions. The position of the $k-$th agent, $x_k(t)$, for the step $t+1$ is obtained according the backward update rule,  
\begin{equation}\label{eq:update_position}
\vec x_k(t+1)=\vec x_k(t)+v_0e^{i\theta_k(t)},
\end{equation}
where the last term is the velocity of the agent and $i$ the imaginay unity. Like the Vicsek's model each component particle interacts with its neighbours lying within a circular region of radius $r_0$. While the resulting interaction forces the particle to move towards its neighbour's averaged motion direction, an additive random perturbation $\eta$ effectively select the resulting direction of motion. Both contributions lead to a behaviour that mimics a difficulty to ``choose'' appropriately a direction of motion. The original model make use of the \textit{angular noise}, equivalent to applying a random rotation in neighbourhood averaged direction $\eta\xi$~\cite{vicsek1995novel,chate2008modeling}. However, we use a variation of the motion direction update proposed by~\citet{gregoire2004onset}:
\begin{equation}\label{eq:vectorial_noise}
\theta_k(t+1)=\arg\left[\sum_{j=1}^{n_k}\left(e^{i\theta_j(t)}+\eta_k e^{i\xi_k(t)}\right)\right].
\end{equation}
This last case is usually known as \textit{vectorial noise} attending for a random vector $\eta_k e^{i\xi_k}$ added to the velocity of each neighbour in the average. In a biological context, this corresponds to a difficulty of the agent to perceive its' environment and is often understood as an "extrinsic" (vectorial) noise. Note that the factor $\eta_k$ in Eq.~(\ref{eq:vectorial_noise}) establishes the heterogeneity of the system: while in the Vicsek model $\eta_k=\eta$, for all $k$, in our model each individual has its own positive noise amplitude, taken from a normally distributed random series around mean noise amplitude $\mu$ with standard deviation $\sigma$.

In the original model the absolute value of averaged velocities, $$\phi(\eta)=\frac{1}{N}\left|\sum_{j=1}^N\vec v_j\right|,$$
corresponds to the order parameter used to identify a gas-like phase from a collective motion phase. When $\eta$ is beyond the transition point, $\eta_c$, $\phi$ is null so the system is in the gas phase, however for $\eta<\eta_c$, $\phi$ increases to one as $\eta$ approaches zero. In case of vectorial noise, a first-order transition occurs between the phases~\cite{gregoire2004onset,solon2015phase}. 
This phase transition can also be detected measuring higher order moments of $\phi$, namely, the susceptibility $$\chi(\eta)=L^2(\langle\phi(\eta)^2\rangle-\langle\phi(\eta)\rangle^2)$$
and its fourth-order Binder cumulant
$$U(\eta)=1-\frac{1}{3}\frac{\langle\phi(\eta)^4\rangle}{\langle\phi(\eta)^2\rangle^2}.$$

\paragraph*{Simulations setup} 
The density, velocity and vision radius for all simulations were set to $\rho=1.0$, $v=0.5$ and $r_0=1.0$, respectively. 
Simulations start with randomly distributed agents pointing in a random directions and a simulation step ends when all the particles have been updated according to rules given by Eq.~(\ref{eq:update_position}) and Eq.~(\ref{eq:vectorial_noise}). 
In a typical realization we periodically assess above mentioned observable long after the system reaches the stationary state and the results were obtained averaging across 100 realizations where for same $\mu$ and $\sigma$ each realization has a different $\{\eta_k\}$ set. 
We must stress that all results are presented as function of $\mu$, the Gaussian pseudo-random number generator input value, that is understood as a control parameter of the system. 

Aware that finite-size effects might smooth discontinuous nature of phase transition, we chose $L\geq L^{*}=64$ as suggested in \cite{chate2008collective}. 
For selected density and velocity parameters, systems beyond $L^{*}$ are expected to behave in the asymptotic regime free of any finite-size effects that might affect the discontinuity of the transition~\cite{chate2008collective}. 
Therefore, we have considered systems larger than $N=4096$ and up to $N=131072$ particles. 

\begin{figure}
\begin{center}
\includegraphics[width=0.33\textwidth]{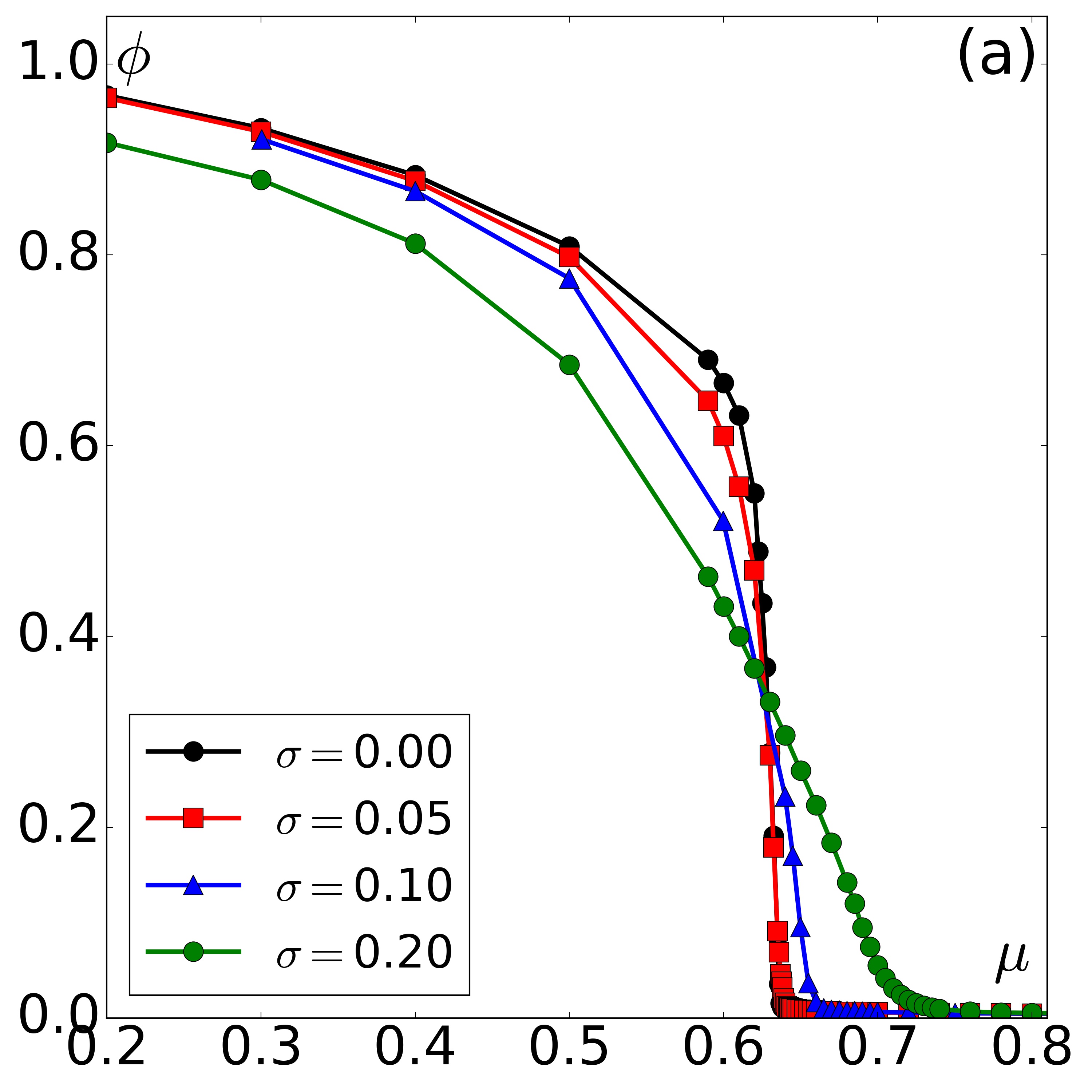}
\includegraphics[width=0.33\textwidth]{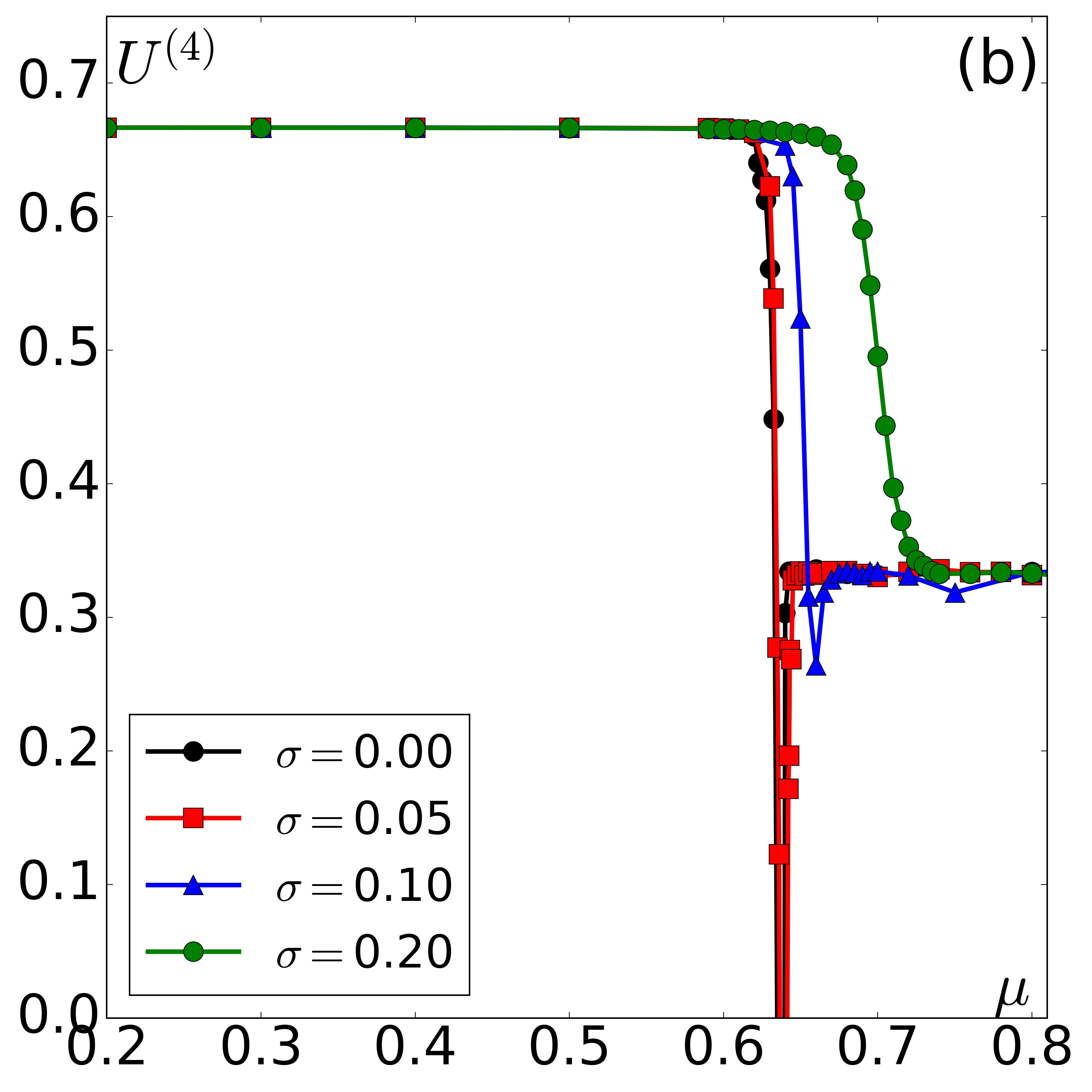}
\\
\includegraphics[width=0.33\textwidth]{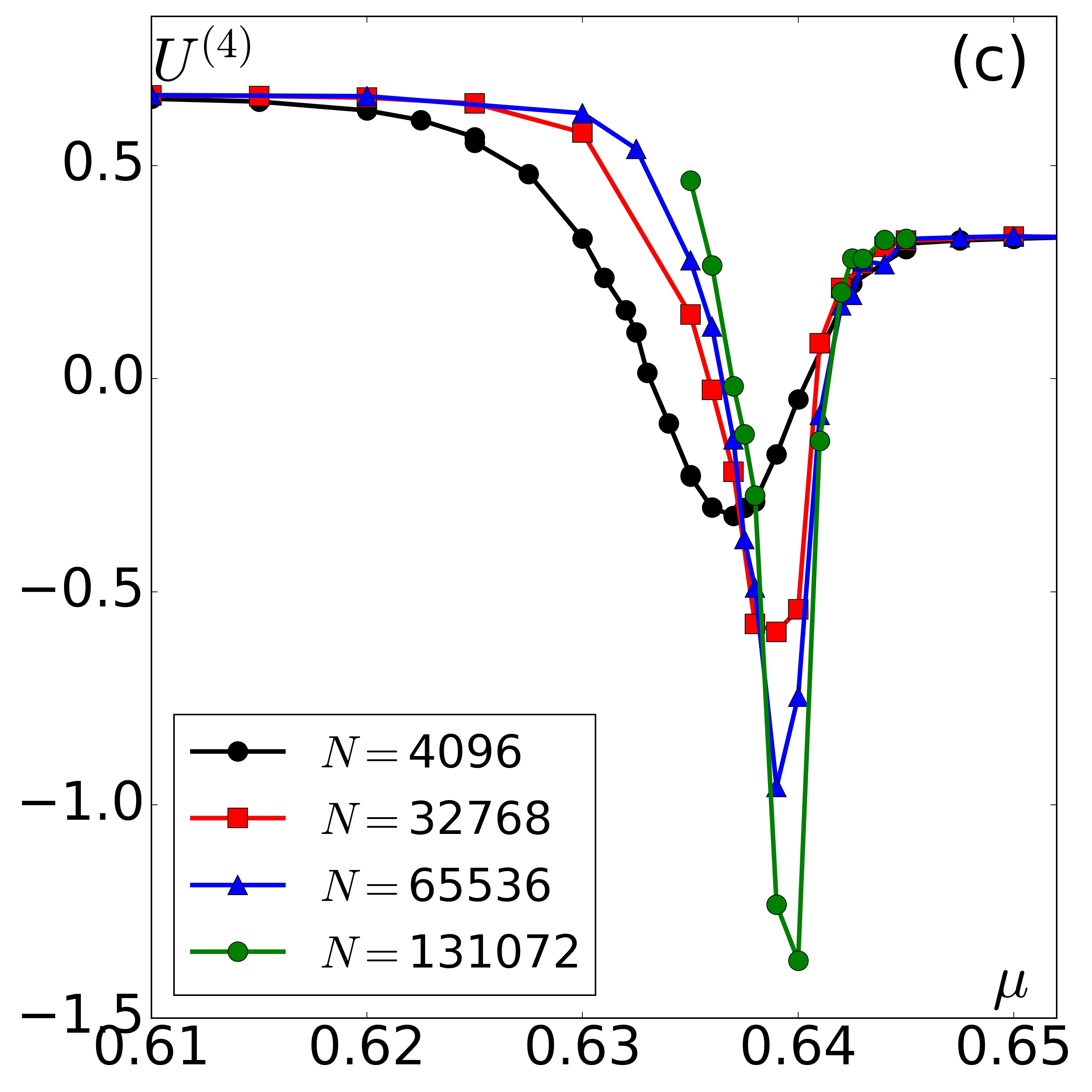}
\includegraphics[width=0.33\textwidth]{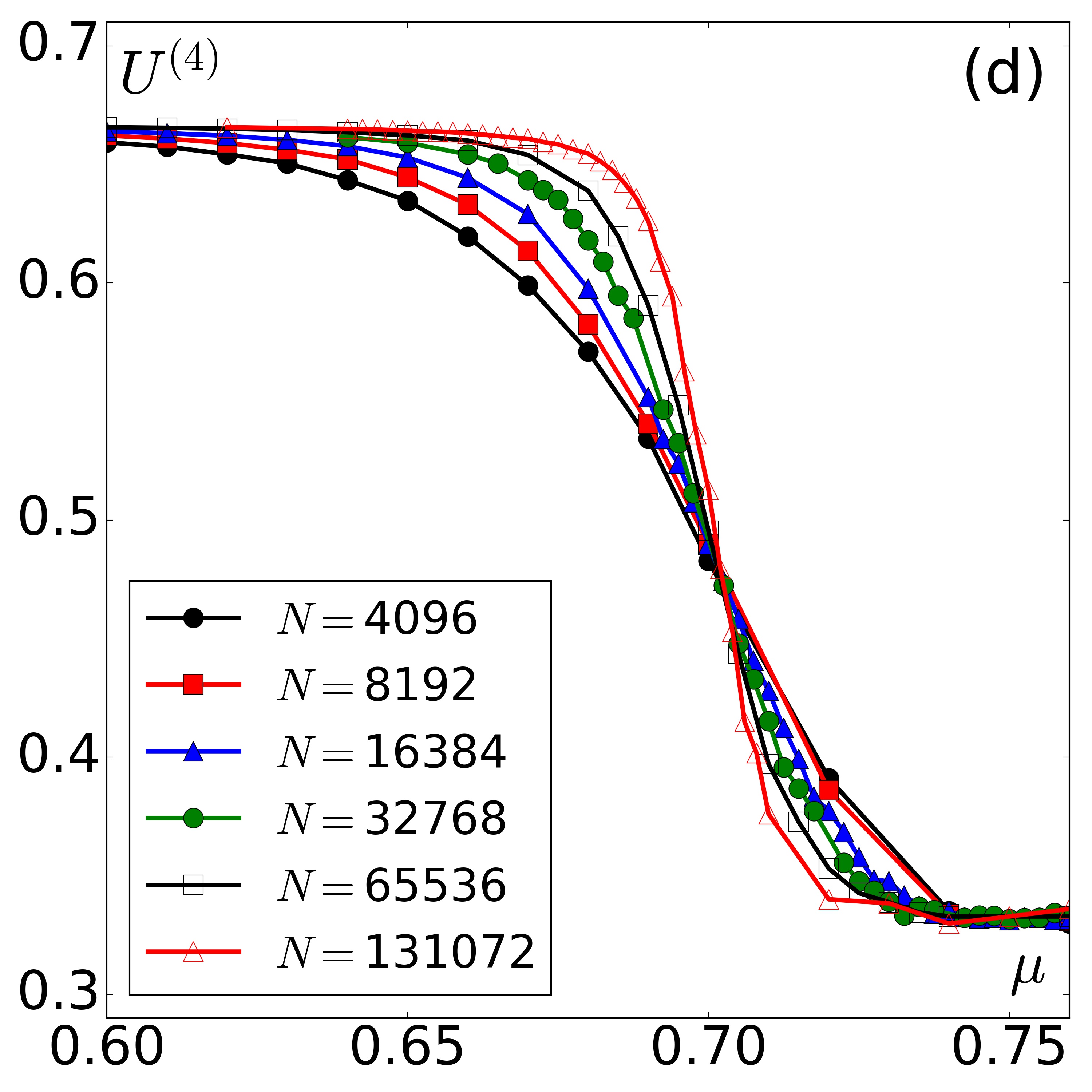}
\end{center}
\caption{Plots of averaged velocity~(a) and the corresponding Binder cumulant~(b) as a function of mean noise $\mu$ for systems of $N=65536$ particles and different $\sigma$ detailed in the legends. Panels~(c) and~(d) shoe Binder cumulant as a function of mean noise amplitude for different sizes setting heterogeneity to $\sigma=0.05$ and $0.20$, respectively.}\label{op_simu}
\end{figure}

\begin{figure*}
    \begin{tabular}{cc}
        \begin{tabular}{c}
            \includegraphics[width=0.275\textwidth]{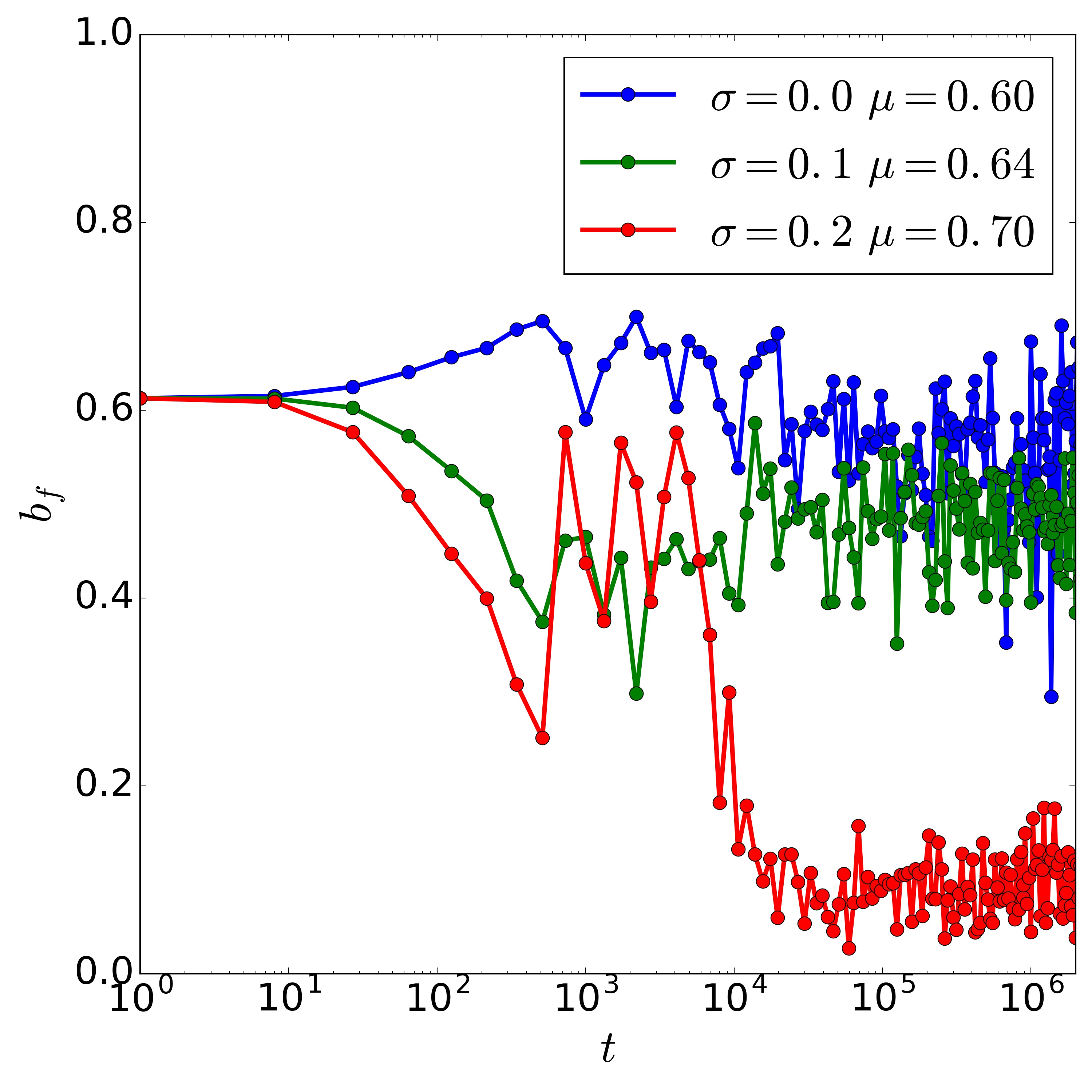}
        \end{tabular}
          &  
        \begin{tabular}{ccc}
            \includegraphics[width=0.15\textwidth]{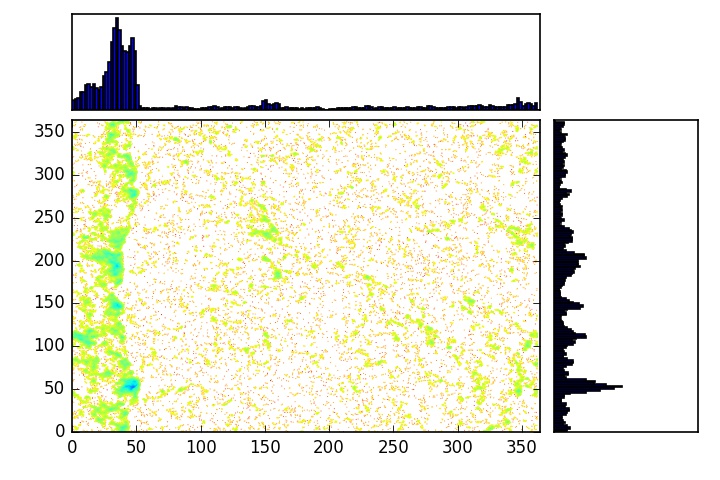} &  
            \includegraphics[width=0.15\textwidth]{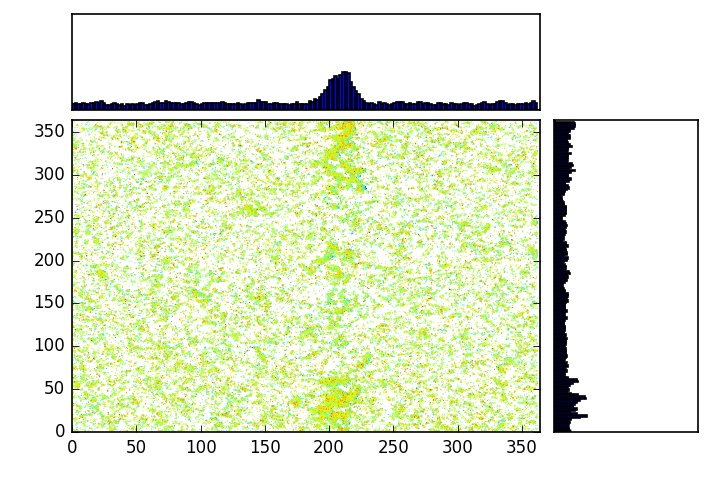} &  
            \includegraphics[width=0.15\textwidth]{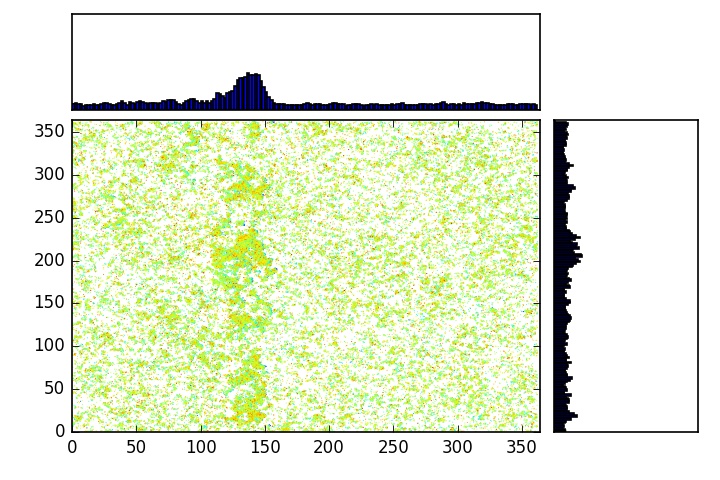}
            \\
            \includegraphics[width=0.15\textwidth]{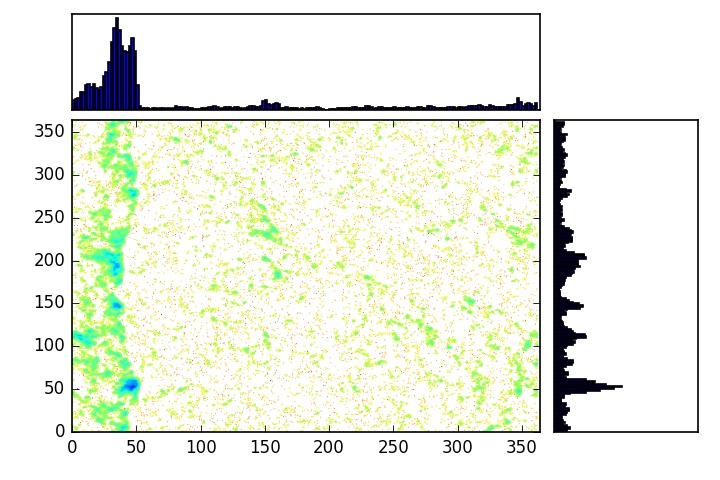} &
            \includegraphics[width=0.15\textwidth]{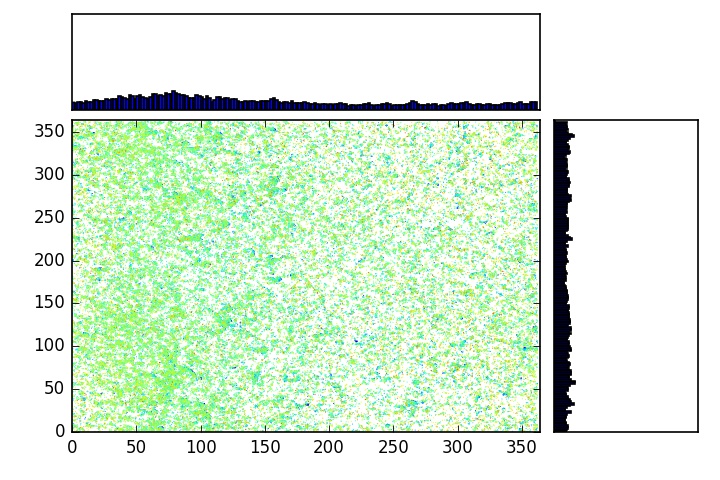} &
            \includegraphics[width=0.15\textwidth]{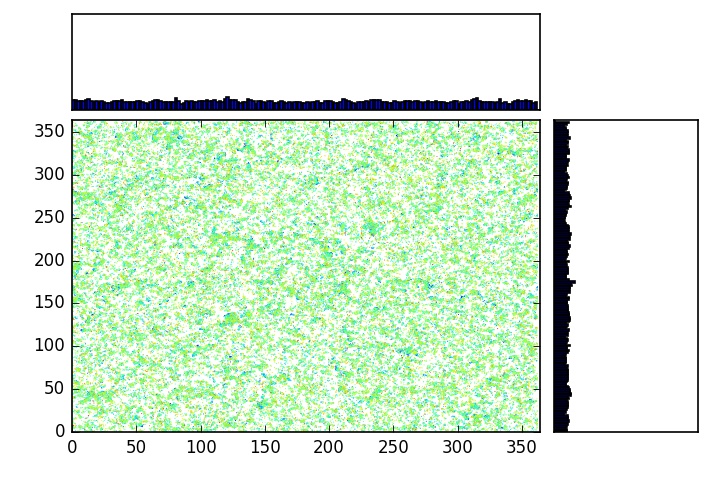} 
        \end{tabular}
    \end{tabular}
\caption{Left panel, time evolution of $b_f$ when the initial configuration is a band. Where a first-order transition occurs ($\sigma=0.0$ and $\sigma=0.1$), the parameter is close to its initial value implying that the band is stable and endures as expected. For $\sigma=0.20$, a continuous phase transition scenario, there is drop in $b_f$ indicating that and initially imposed band cannot hold after a finite amount of steps. The right panel shows how the bands evolution for $\sigma=0.1$, $\mu=0.64$ (top) and $\sigma=0.2$, $\mu=0.70$. In the first case the band is stable in a period of 4~M steps (upper-right panel) however in the same span of time the band completely disappears for $\sigma=0.20$ (lower-right panel). Blue colored dots represent lower noise particles whereas red dots, higher noise particles.}\label{snaps}
\end{figure*} 

\section{Results}\label{results}
Figure~\ref{op_simu}~(a) shows the plots of the order parameter $\phi$ as a function of the noise amplitude mean value, $\mu$. 
Increasing $\sigma$ leads to a rounding and a shift of the transition originally seen for $\sigma=0$ at $\mu_c\approx0.62$. 
Due to heterogeneity, there is a drop in $\phi$ even when $\mu<\mu_c$. 
For these values of $\mu$ there are high-noise agents prone to disband from flocks and thus increasing the system's disorder. 
On the other hand, beyond $\eta_c$ there is a non-negligible interval with non-zero velocity for $\sigma\neq0$. 
In this case, the presence of low-noise agents liable to form flocks increases system's average velocity, as was also observed for binary mixtures~\cite{ariel2015order}. 
The corresponding Binder cumulant assessments are plotted in Figure~\ref{op_simu}~(b).
Here we can observe that for $\sigma=0$ the curve exhibits a negative sharp peak indicating the first-order nature of the transition~\cite{chate2008collective,binder1997applications}.
However, slightly increasing sample's heterogeneity makes the peak to shrink ($\sigma=0.05$ and $\sigma=0.10$) and eventually disappear for $\sigma=0.20$. 
In this last case, the transition from low to high-noise regimes is smooth and monotonic. 
To rule out finite-size effects screening first-order transition deportment we consider a finite-size scaling approach. 
In Figure~\ref{op_simu}~(c) and~(d) it can be seen that Binder cumulants for $\sigma=0.05$ and $\sigma=0.20$, respectively, clearly show two different behaviours. 
For the low-heterogeneity case the bigger the system the sharpen the peak, implying that a first-order phase transition takes place at thermodynamic limit, as in the homogeneous case. 
However for $\sigma=0.20$, Binder cumulants corresponding to different sizes, cross each other close to $\mu_c=0.70$ strongly suggesting continuous phase transition at the crossing point~\cite{ginelli2010relevance}.

\begin{figure}
\begin{center}
\includegraphics[width=0.3\textwidth]{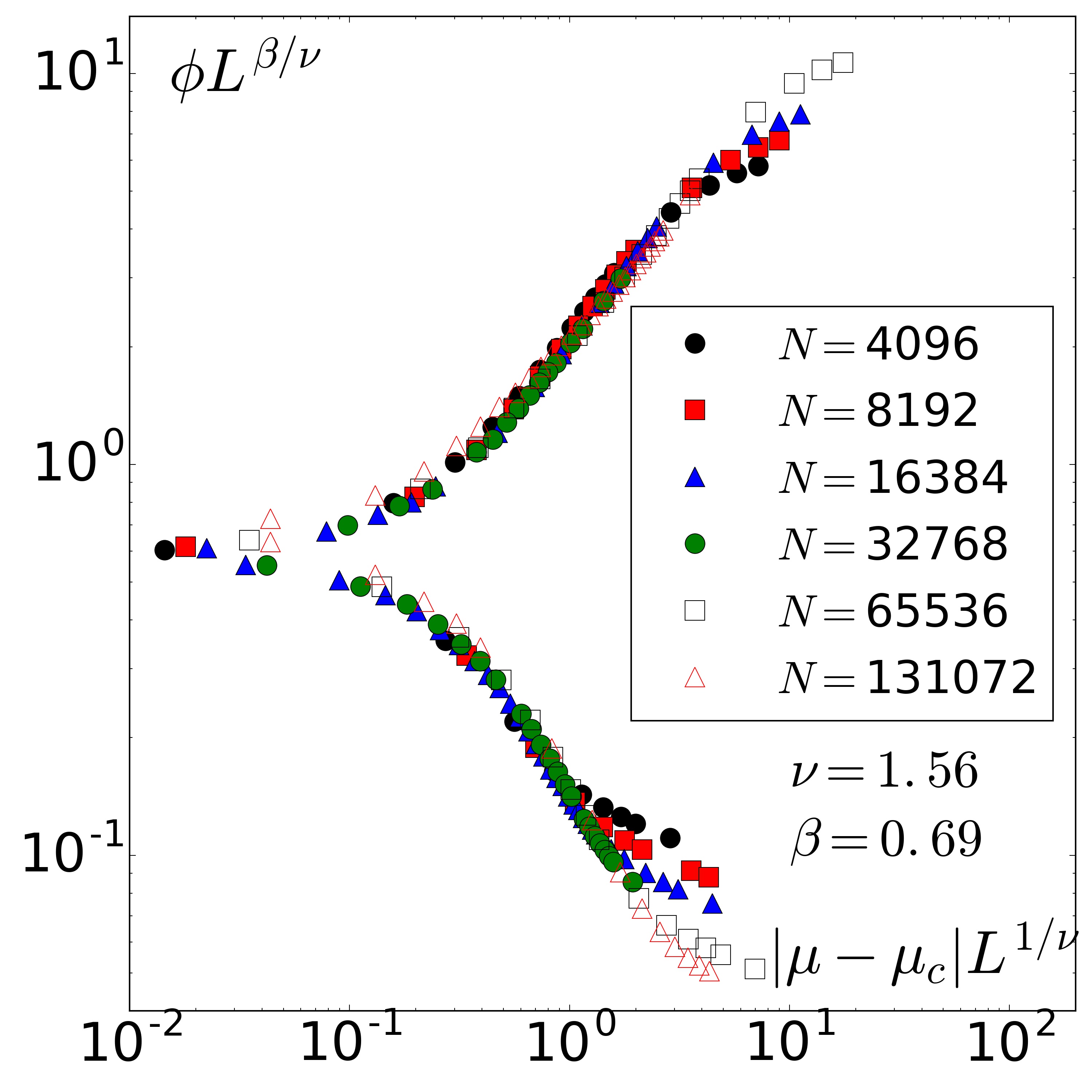}
\includegraphics[width=0.3\textwidth]{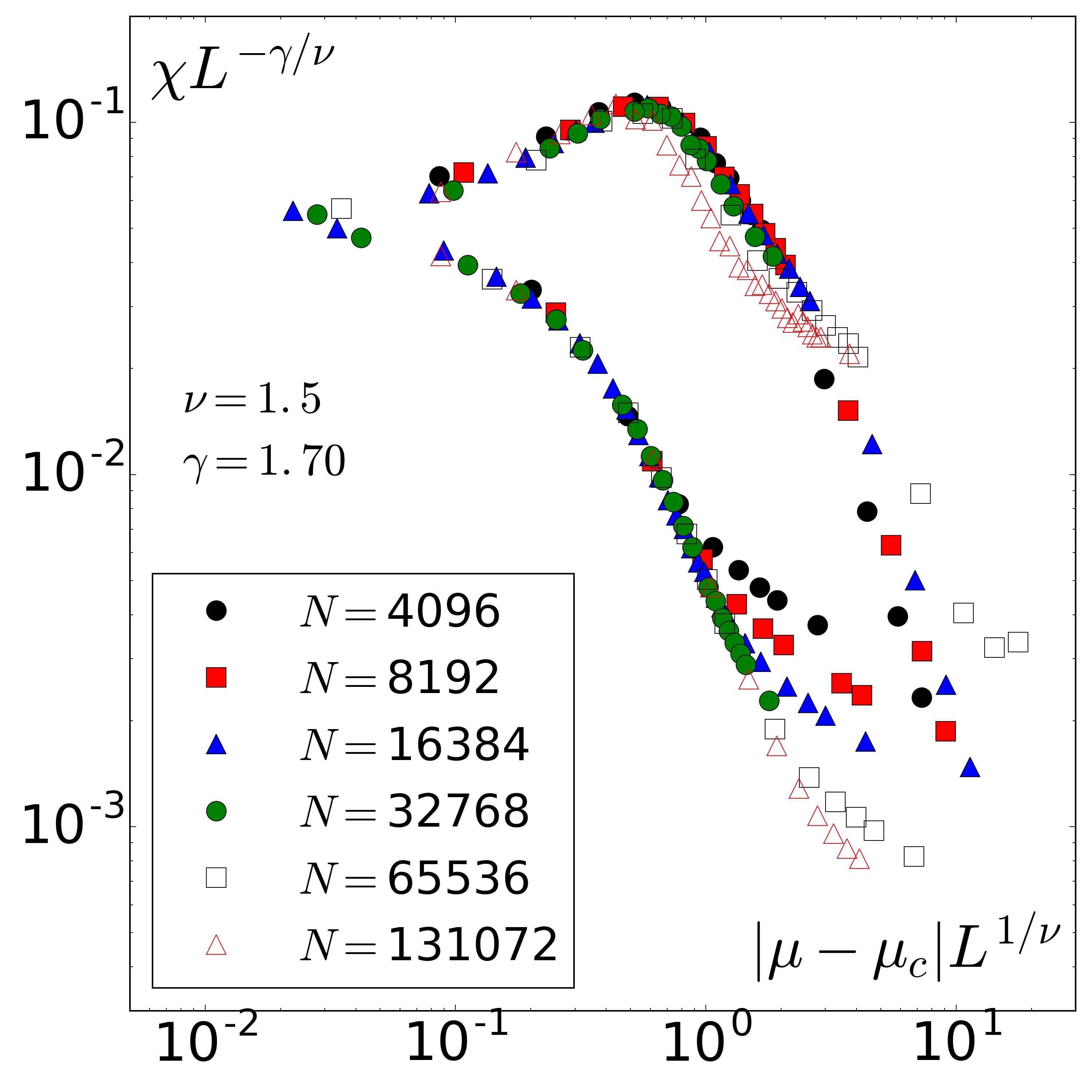}
\includegraphics[width=0.3\textwidth]{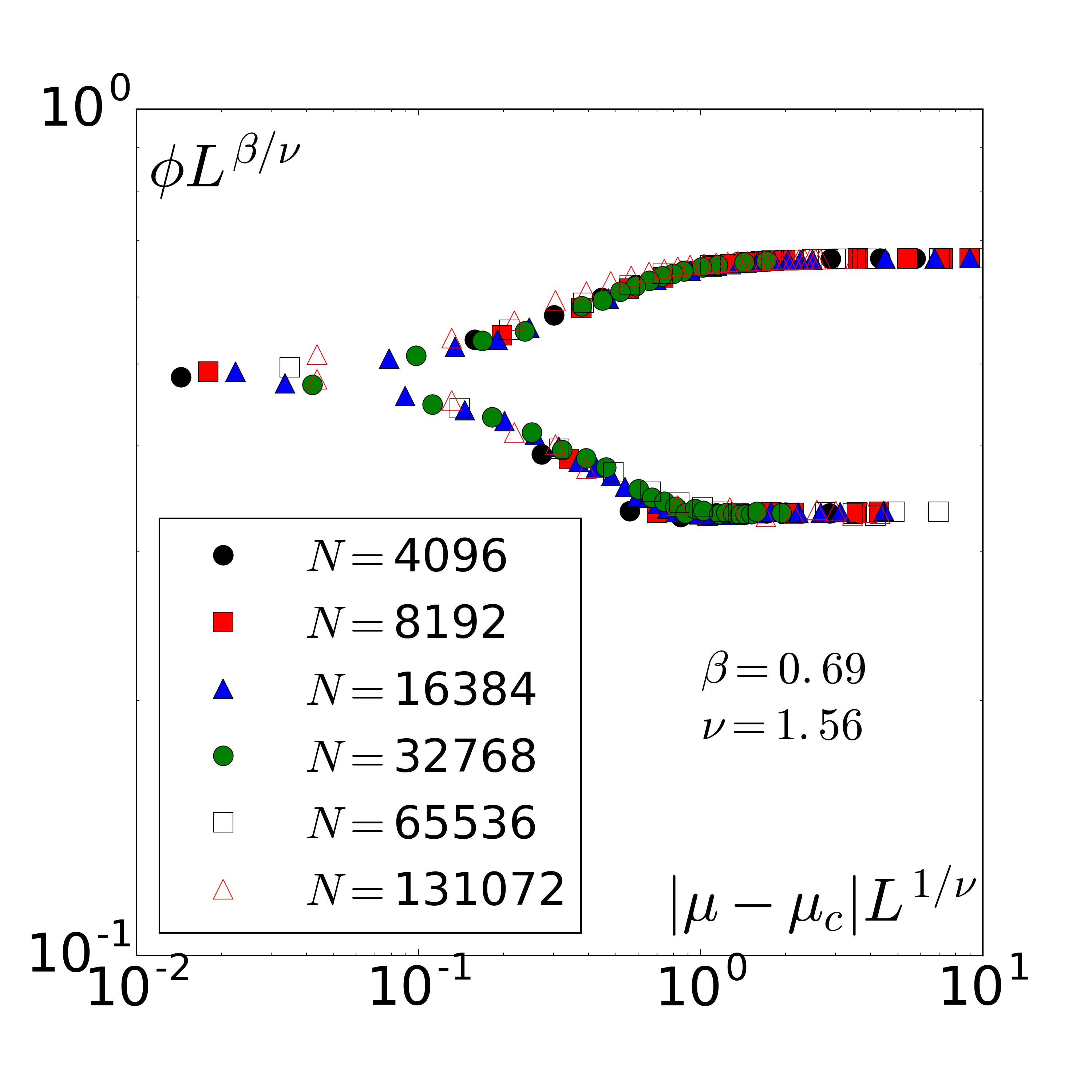}
\end{center}
\caption{Scaling log-log plots of the order parameter, susceptibility on the right and left panels, respectively, for $sigma=0.20$. Satisfactory data collapses were obtained using the exponents detailed in the plots and $\langle\mu\rangle_c=0.70$.}\label{collapse_sigma020}
\end{figure}

At the onset of collective motion the coexistence of high-density ordered bands with low-density disordered background is an undeniable signature of a first-order transition \cite{solon2015phase,gregoire2004onset,chate2008modeling}. 
In order to characterize the presence of bands in the system we define a new parameter related to the linear particle density. 
When a band moves along one of the axes of the box, e.g. the $x$-axis, the particle density along this direction is sharper than the $y$-axis density. 
While in the $y$-direction the particle density is approximately constant, in the $x$-direction there is a peak indicating the location of a band. Then, the standard deviation of the latter density, $\sigma_{\rho_x}$, is smaller than corresponding to the former, $\sigma_{\rho_y}$. 
If there is no band, both densities have a similar flatness and the standard deviations are approximately the same. 
Defining band factor index, 
\begin{equation}
b_f=1-\frac{\min(\sigma_x,\sigma_y)}{\max(\sigma_x,\sigma_y)},
\end{equation}
we have a parameter that will be close to zero if no band is present in the system and will approach to the unity as a band increases its sharpness. 
Measuring the time evolution of $b_f$ we can study the stability of bands for systems with different heterogeneity close to their transition points. 
As initial configuration we set a band moving along the $x$-axis direction, which was obtained for the Vicsek model with vectorial at the band phase. 
For $\sigma=0$ and $\sigma=0.1$ the parameter maintain a value close to the initial one indicating that the band travels through the box almost with no modification (see left panel of Figure~\ref{snaps}). 
On the other hand, for $\sigma=0.2$ the parameter shows a significant drop, similar to the one observed in the gas-like phase (not shown here). 
This test shows us that the heterogeneity strikes the bands stability allowing the system to bear a continuous phase transition. 
Furthermore, on a visual inspection of system's snapshots as the one shown in right panel of Figure~\ref{snaps}, we can see that for low heterogeneity, $\sigma=0.10$, there is a stable band traveling through the box (upper-right panel) but for higher heterogeneity, $\sigma=0.20$, the band progressively disappears ruling out any coexistence context (lower-right panel). 

\begin{figure}
\begin{center}
\includegraphics[width=0.45\textwidth]{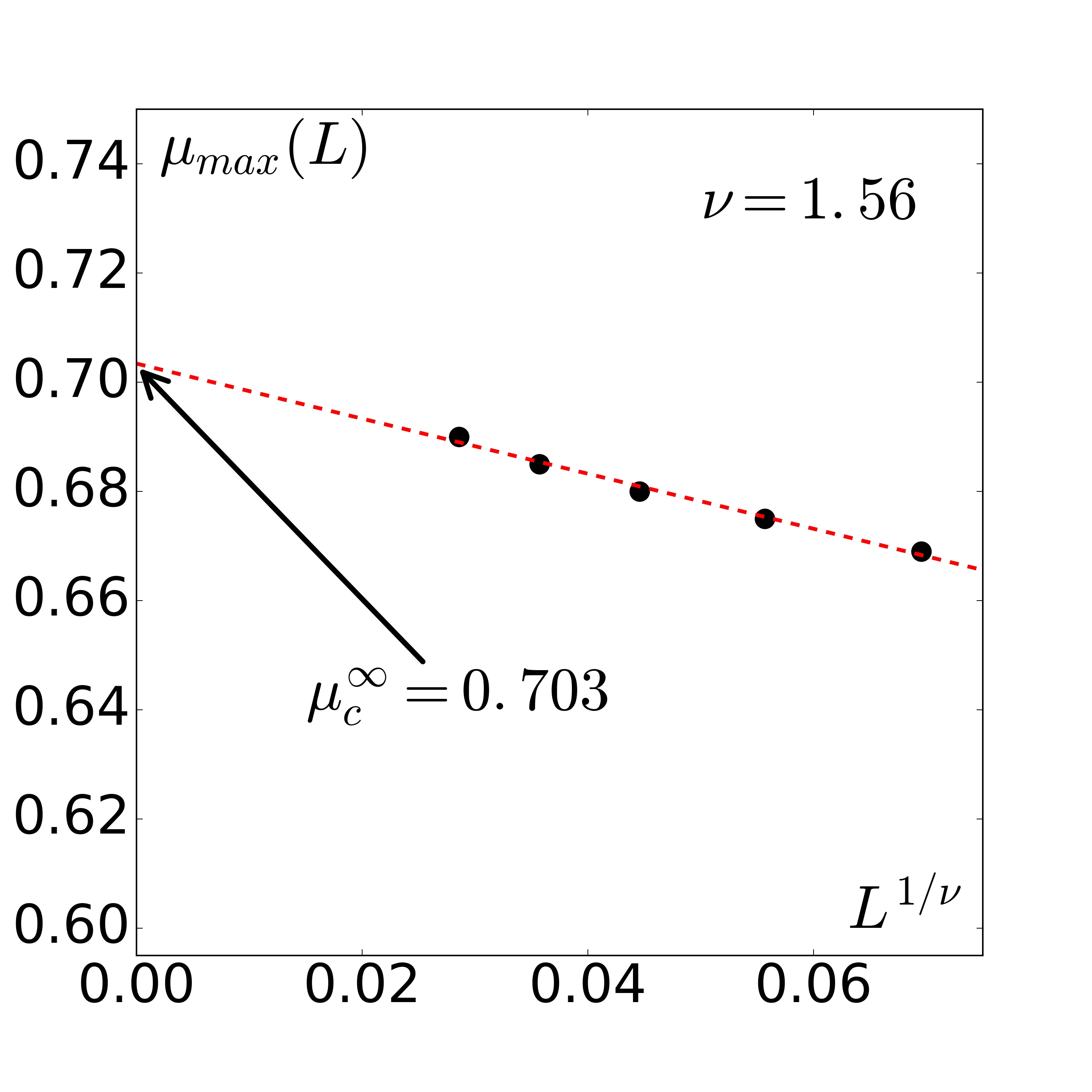}
\includegraphics[width=0.45\textwidth]{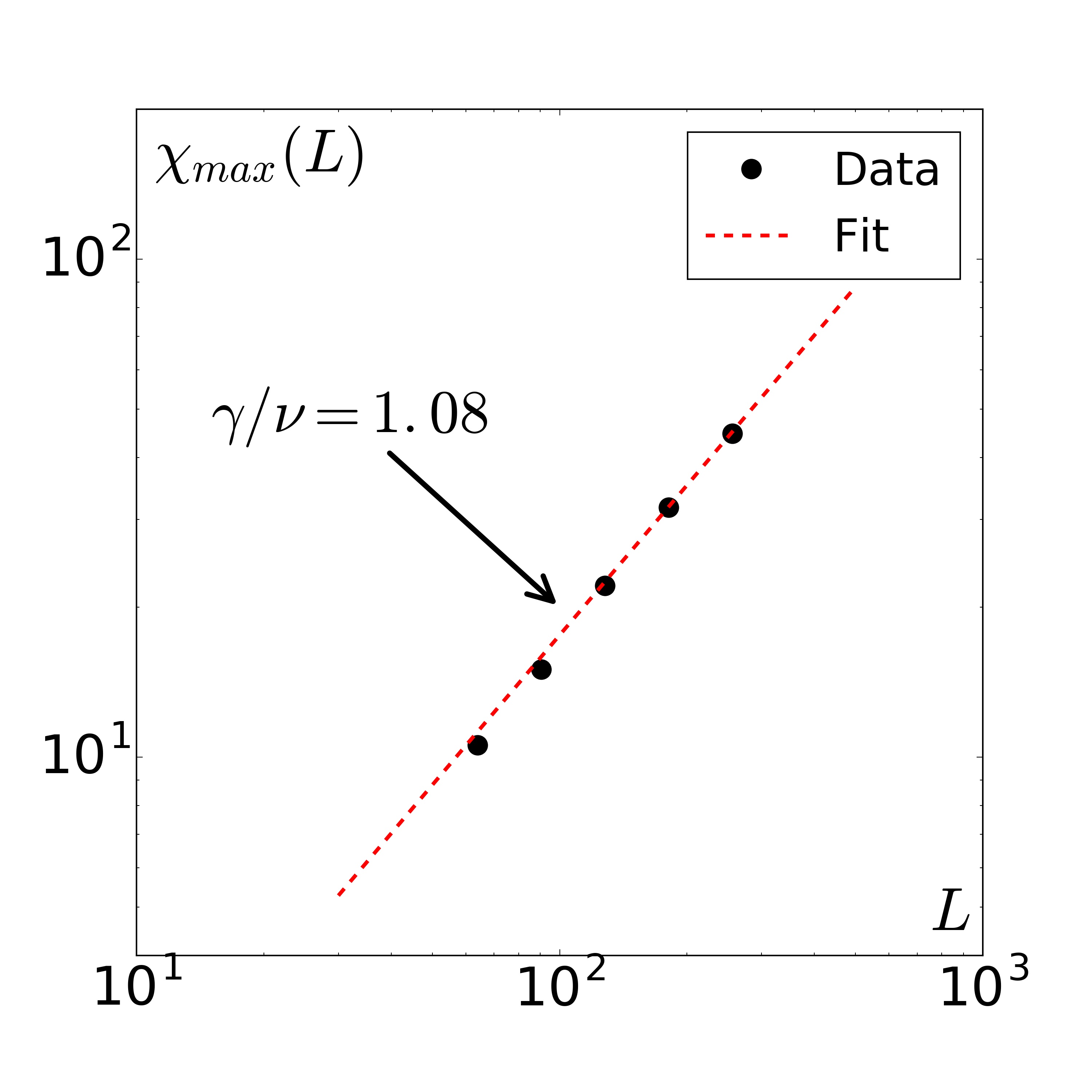}
\end{center}
\caption{Scaling plots of the susceptibility. In the left panel the scaling plot for susceptibility peak position as a function of $L^{1/\nu}$. The lineal regression, dashed line, fits the data very well. The extrapolation to thermodynamic limit gives expected mean-noise value where the phase transition occurs, $\langle\mu_c\rangle=0.70$ in accordance to the value where  Binder cumulants cross each other, see Figure~\ref{op_simu}~(d). Right panel shows log-log scaling plot of susceptibility height as a function of system characteristic length. A power-law with slope $\gamma/\nu=1.08$, dotted dashed line, fits very well the data. }\label{suscep_peaks}
\end{figure}

Let's now focus on $\sigma=0.20$ case and estimate the corresponding exponents characterizing the phase transition by means of standard finite-size scaling techniques~\cite{chate2008modeling,ginelli2010relevance}. 
We make use of an averaged velocity scaling ansatz accepted in the literature~\cite{vicsek2012collective}, $$\phi(\mu,L)=L^{-\beta/\nu}\tilde\phi(|\mu-\mu_c|L^{1/\nu}),$$
where $L$ stands for system size, $\tilde\phi$ a scaling function, $\beta$ and $\nu$ scaling exponents corresponding to order parameter and correlation length, respectively. 
In a similar way, a scaling law for the susceptibility holds $$\chi(\mu,L)=L^{\gamma/\nu}\tilde\chi(|\mu-\mu_c|L^{1/\nu}),$$
with $\gamma$ the susceptibility exponent and $\tilde\chi$ a scaling function and the Binder cumulant
$$U(\mu,L)=\tilde U(|\mu-\mu_c|L^{1/\nu}).$$
Best data collapse for $\phi$, $\chi$ and $U$ scaling laws are shown in the left, middle and right panels of Figure~\ref{collapse_sigma020}, respectively. 
For these collapses we obtained $\beta=0.69$, $\gamma=1.7$ and $\nu=1.56$ that, in addition, hold for hyper-scaling relation $d\nu-2\beta-\gamma=0$ with good accuracy for system dimension $d=2$. 
We have also acknowledged consistency checks of the above estimation using additional scaling relationships on maximal susceptibility and peak position for different system sizes, $L$. 
Due to finite-size effects, susceptibility peak position shifts respect from expected at thermodynamic limit ($\eta_c^\infty$) according to $$\mu_c^{eff}(L)=\mu_c^\infty+AL^{1/\nu},$$ where $A$ is a constant~\cite{ginelli2010relevance}.
In top panel of Figure~\ref{suscep_peaks}, $\mu^{eff}_c$ is plotted as a function of $L^{1/\nu}$. 
The extrapolation of the effective critical noise to the thermodynamic limit, $L^{1/\nu}\rightarrow0$, gives $\eta_c^\infty=0.703\pm0.005$ for $\nu=1.56$, i.e., the critical noise amplitude in the thermodynamic limit. 
This value is in very good accordance with the crossing point seen in Figure~\ref{op_simu}~(d). 
On the other hand when plotting the height of the susceptibility peaks, $\chi_{max}(L)$, as a function of $L$ we observe a power-law behaviour where slope $\gamma/\nu=1.08$ fits data very well. 
This is also consistent with scaling law on maximal values of the susceptibility given by     $\chi_{max}(L)\propto L^{\gamma/\nu}$~\cite{ginelli2010relevance}.

\begin{figure}
\begin{center}
\includegraphics[width=0.45\textwidth]{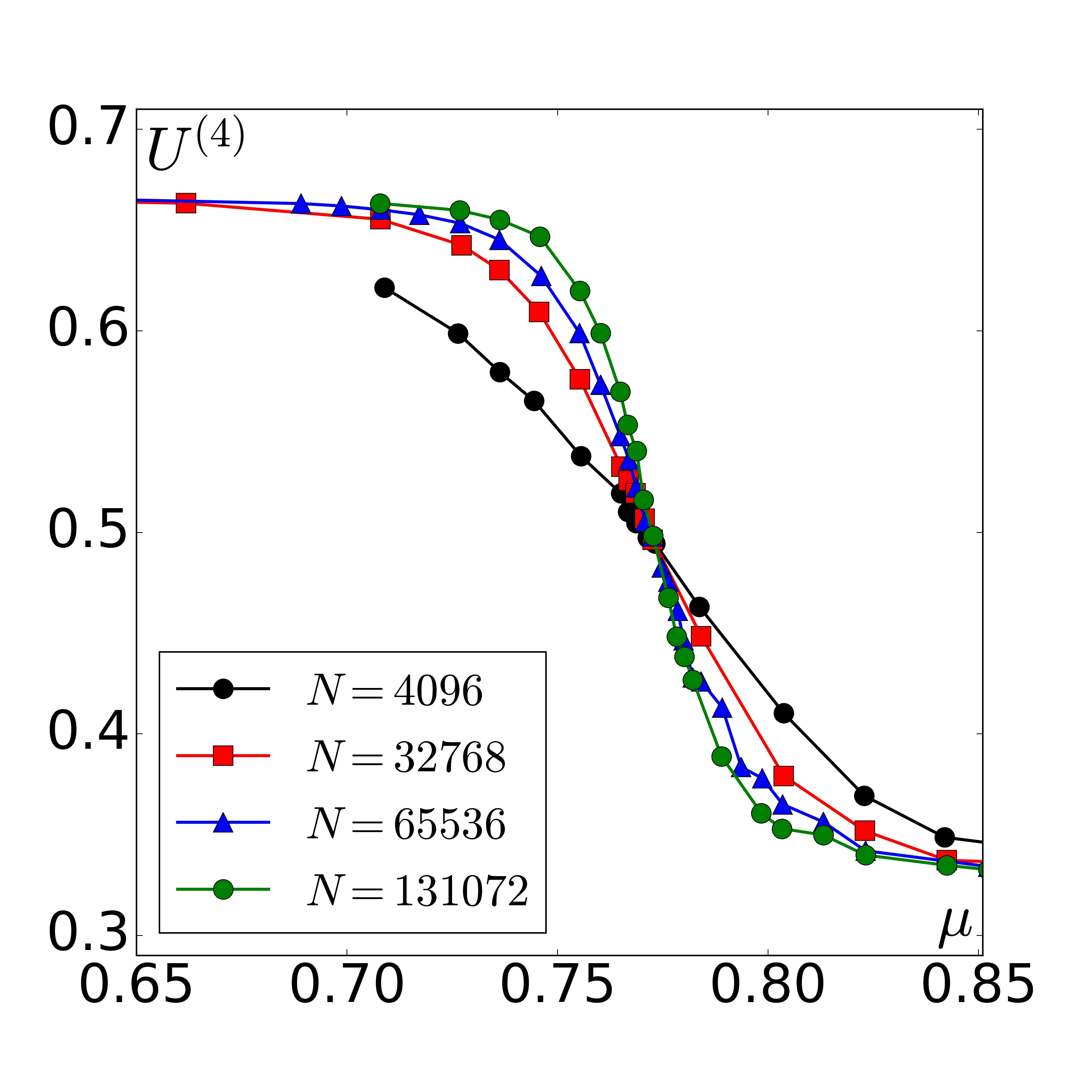}
\includegraphics[width=0.45\textwidth]{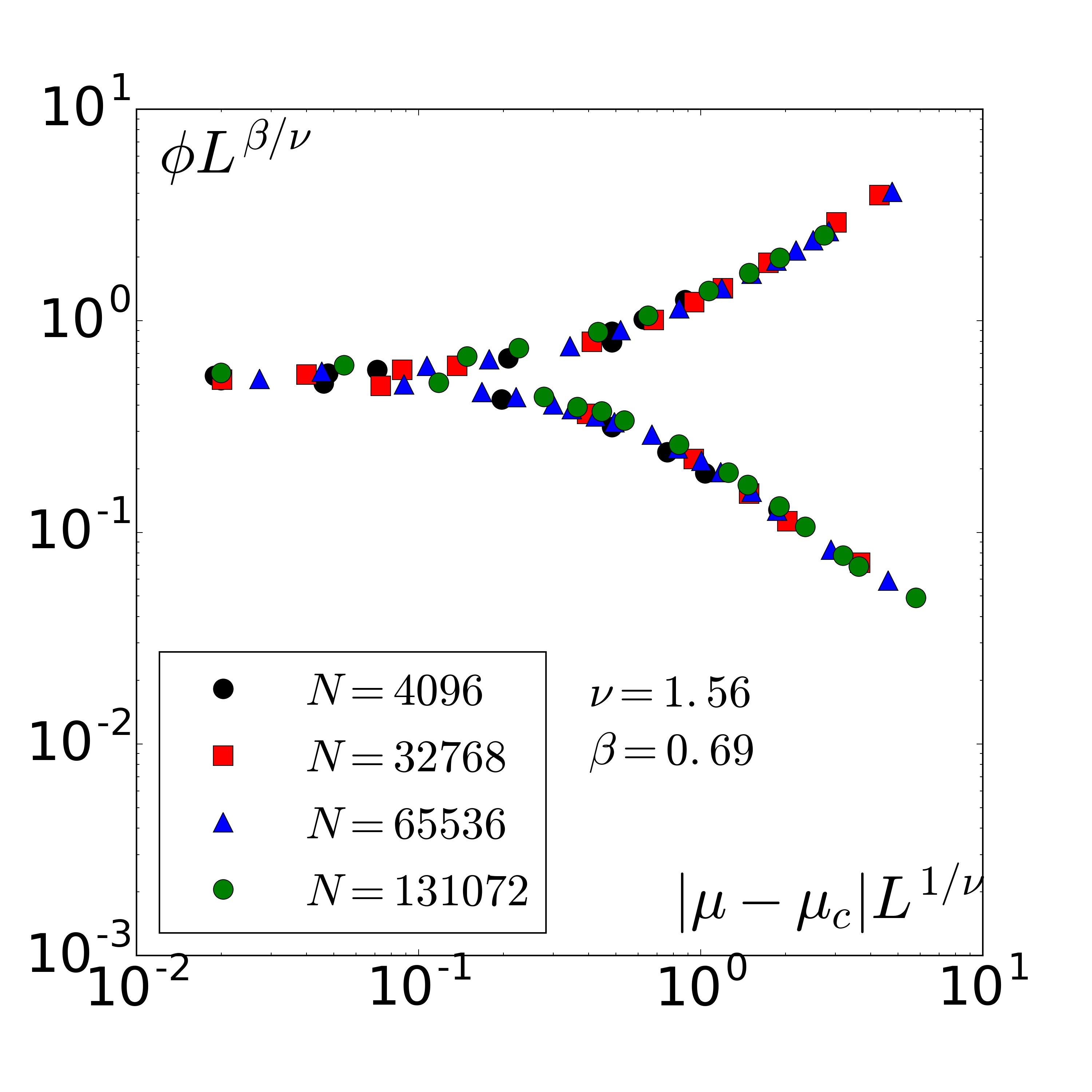}
\end{center}
\caption{On the left panel we show plots of the Binder cumulant for different sizes as a function of noise mean value for $\sigma=0.30$. No peak is observed but all curves crossing each other in one point, as it happens in a continuous phase transition. On the right panel we present a log-log scaling plot of the order parameter for $\sigma=0.30$. Using previously obtained exponents for $\sigma=0.20$ gives a remarkably good data collapse.}\label{plots_sigma030}
\end{figure}

We have also considered normally distributed noise with standard deviation set to $\sigma=0.30$. 
As can be seen in the left panel of Figure~\ref{plots_sigma030}, Binder cumulants exhibit the same behaviour as observed for $\sigma=0.20$ although there is a clear shift in critical noise value where the transition takes place. 
This shift is related to the fact that for larger $\sigma$ there are more agents with very low noise (below critical value) not keen on changing their movement direction, thus a larger $\mu$ is needed to reach a gas-like behaviour. 
One might then ask whether the exponents driving the transition are the same as in $\sigma=0.20$ or there is a crossover depending on distribution standard deviation. 
We have considered the same scaling-law previously defined with the exponents obtained for $\sigma=0.20$ and found a very good data collapse, shown in the right panel of Figure~\ref{plots_sigma030}.

\begin{figure}
\begin{center}
\includegraphics[width=0.75\textwidth]{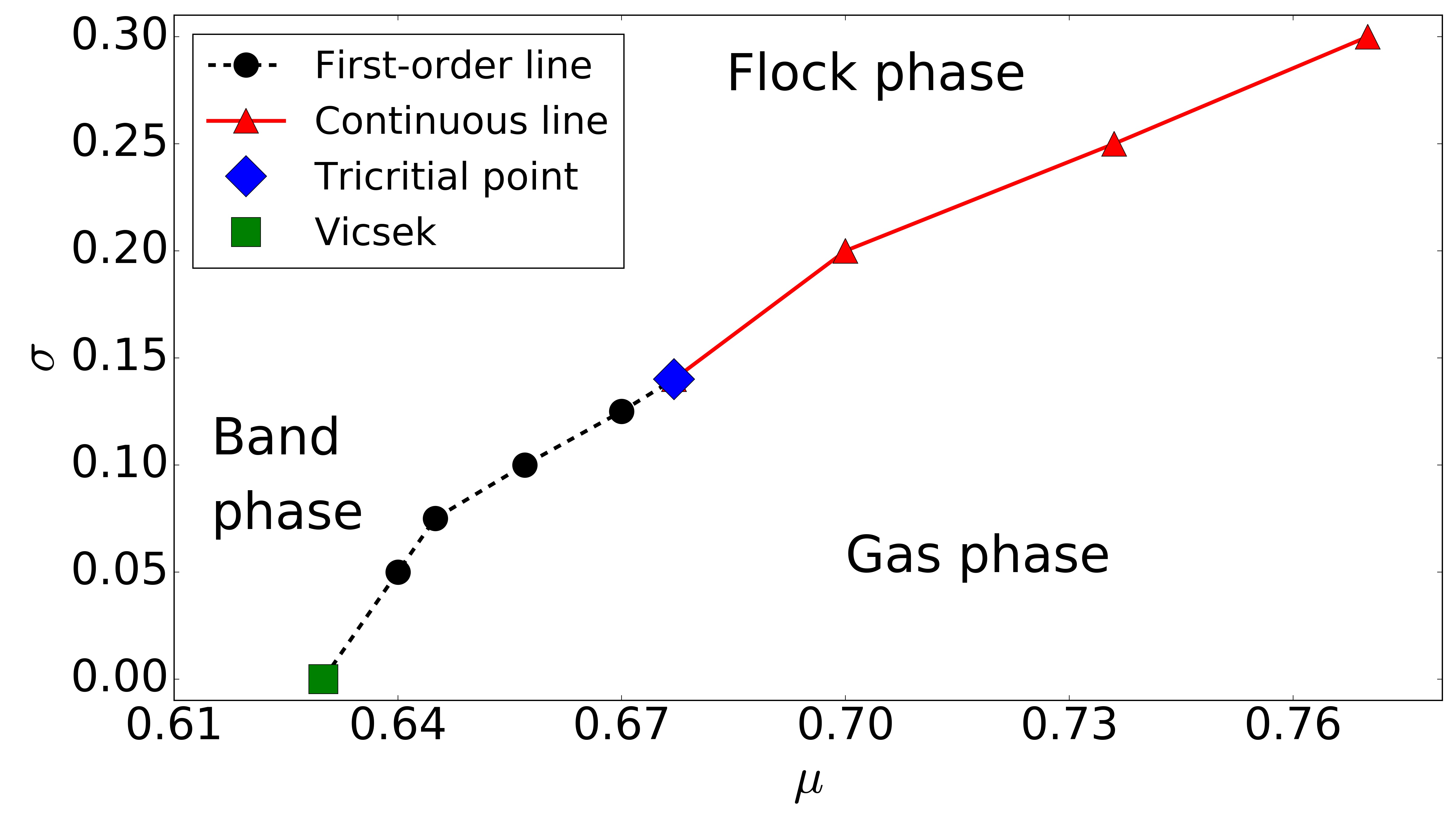}
\end{center}
\caption{Phase diagram of the Vicsek model with heterogeneity. The dashed line with full circles shows the first-order transition line between the gas-like and liquid phases. The solid line with full triangles corresponds to the continuous phase transitions for large values of heterogeneity. The full square shows the location of the already known first-order phase transition for the Vicsek model with vectorial noise and the approximate location where both lines of transitions meet, the tricritical point  $\sigma_{tri}\approx0.15$, is marked with a full diamond.}\label{phase_diagram}
\end{figure}

Exploring values of heterogeneity up to $\sigma=0.3$ we have found a first-order transition line for $\sigma<\sigma_{tri}\approx0.15$, see Figure~\ref{phase_diagram}. 
In this region the collective motion manifest itself as bands that coexists with low-density regions of agents. 
Since the behaviour of the system is the same as for the already known homogeneous case, we denominate this region the \textit{band phase}. 
Beyond $\sigma_{tri}$ there is a line of continuous phase transitions driven by the same critical exponents. 
In this regime as gas-like swarm of agents get together in flocks of different sizes when the noise is decreased and no band is observed thus we differentiated it from the band phase identifying as \textit{flock phase}. 

\section{Discussion}\label{conclusions}
Considering samples of self-propelled particles with noises normally distributed around a mean value $\mu$, we used standard deviation, $\sigma$, as a proxy to heterogeneity. 
As a result we observed that collective behaviour of the VM is strongly affected. 
Low-heterogeneity populations behave as a homogeneous ones where a first-order transition between a gas-like uncorrelated state and a collective motion state (bands) is observed, a scenario that is well understood as a liquid-gas microphase separation~\cite{solon2015phase,chate2008collective}.
However when increasing the heterogeneity the system undergoes a continuous phase transition. 
For this regime, we have acquired satisfactory collapses of the measured observable and finite-size techniques strongly validate the obtained critical exponents. 
We have also found that the bands become unstable for large values of heterogeneity, in contrast with the low-heterogeneity case, where they persist. 
The formation of percolating high density traveling bands in the Vicsek model is strongly related to the use of boundary conditions in the treatment (numerical or analytical) of the model~\cite{nagy2007new,baglietto2009nature,chate2008modeling}. 
The stabilization of these bands occurs for values of the noise inside the ordered phase and close to the coexistence point and requires the coherent motion of the particles forming them. 
When the heterogeneity increases, the fraction of particles corresponding to band-regime noises in not enough to assemble a flock that percolate to form band. 
In fact, agents with different noise amplitudes travel with different effective drift speeds: those with lower noise amplitude move in a ballistic fashion but those with larger noise amplitude travel in a more diffusive way, reducing their effective drift speed.

Since the standard Vicsek Model has been frequently compared with the equilibrium XY Model for magnetic systems, it could be interesting to compare the results found in this paper with what expected for equilibrium magnetic systems with quenched noise. 
As we have already stated, quenched disorder (heterogeneity) is responsible of rounding the abrupt transition observed in some disordered systems. 
In particular, there is no continuous symmetry breaking for $d\leq4$~\cite{imry1979influence,aizenman1990rounding}.
Although these studies were mainly developed for random-field models with regular topologies in thermal equilibrium, similar results were found for two-dimensional systems out of equilibrium with absorbing states, where no first-order phase transition was observed when quenched disorder was introduced~\cite{martin2014quenched}. 
As the authors claim, this means that a heterogeneous non-equilibrium system will not exhibit coexistence of phases and will move from one phase to the other in a smooth fashion instead in an abrupt one. 
However, this is not the scenario we have found in this study: for small amounts of heterogeneity, first-order phase transition still persists and only for sufficiently large noises a continuous phase transition is observed.

We would like to stress some similarities and differences between our proposal and topological models~\cite{ginelli2010relevance,barberis2014evidence,cavagna2010scale} which also shows a continuous phase transition. 
In a sense, the latter can be understood as a heterogeneous system since a given agent interacts with it's neighbours independently of its geometrical distance. 
This leads to a distribution of effective radii of vision setting up a geometrical heterogeneity.
Nevertheless, one must be aware that in topological models the heterogeneity is dynamic due to the fact that an agent will adjust it's effective radius of vision from one step to another, i.e., the agent will set it's vision radius according to it's surroundings.
The model considered in this work, however, introduce a quenched noise: agents don't change how their perceive their neighborhood and each one will behave different for identical conditions, in a similar manner observed by~\citet{maye2007order}. 
In this way the model mimics the variability of individuals traits present in biological systems. 

As a final remark we claim perception heterogeneity degree as a possible mechanism to conciliate the existence of both first and second order phase transition flocking systems in nature.

\printbibliography

\end{document}